\documentclass[preprint,showpacs,preprintnumbers,amsmath,amssymb,nofootinbib]{revtex4-1}

\usepackage{graphicx}
\usepackage{dcolumn}
\usepackage{amssymb}
\usepackage{mathrsfs}
\usepackage{amsmath}
\usepackage{epsfig}
\usepackage{hhline}
\usepackage[utf8]{inputenc}
\usepackage{color}
\usepackage{multirow}
\usepackage[T1]{fontenc}
\usepackage{verbatim}
\usepackage{slashed}
\usepackage{hyperref}

\begin{document}

\title{Probing vector-like top partner from same-sign dilepton events at the LHC}
\author{Hang Zhou}
\author{Ning Liu}
\email[Corresponding author: ]{liuning@njnu.edu.cn}
\affiliation{Physics Department and Institute of Theoretical Physics, Nanjing Normal University, Nanjing, 210023, China}

\begin{abstract}
The Standard Model is a successful theory but is lack of a mechanism for neutrino mass generation as well as a solution to the naturalness problem. In the models that are proposed to simultaneously solve the two problems, heavy Majorana neutrinos and top partners are usually predicted to lead to a new decay mode of the top partner mediated by the heavy Majorana neutrinos: $T\to b\,W^{+}\to b\,\ell^{+}\ell^{+}q\bar{q'}$. In this paper, we will study the observability of such a new signature via the pair production process of top partner $pp \to T\bar{T}  \to 2b+ \ell^\pm \ell^\pm+4j$ in a model independent way. By performing Monte-Carlo simulation, we present the $2\sigma$ exclusion limits of the top partner mass and mixing parameters at the HL-LHC. 

{\text{Keywords: }}Vector-like top partner; Majorana neutrino; LHC.
\end{abstract}
\maketitle

\section{Introduction}
\label{sec:intro}
The Standard Model (SM) has been a successful low-energy effective theory in describing microscopic phenomena and was completed by the discovery of the Higgs boson in 2012 at the LHC. However, a theory beyond the SM (BSM) is necessary from both theoretical and experimental points of view, one of which is the so-called naturalness problem. With the mass of the observed scalar ($\sim125$\,GeV) being comparable to the electroweak scale ($\sim10^{2}$\,GeV), the naturalness arguments require some mechanism or symmetries to suppress or cancel out the large quadratic divergence when considering loop corrections from heavy particles, such as the SM top quark, which can lead the Higgs mass up to the Planck scale ($\sim10^{19}$\,GeV) instead of the electroweak one. Many BSM models, such as the little Higgs models~\cite{Perelstein:2003wd,Matsedonskyi:2012ym} and composite Higgs models~\cite{Dugan:1984hq,Kaplan:1991dc,Contino:2006qr,Contino:2006nn}, have been proposed to solve this problem by introducing a spontaneously broken global symmetry, leading the Higgs boson to be a pseudo Goldstone boson. 
In these BSM models, vector-like top partners (VLT), usually referred to as $T$, are predicted. VLTs are color-triplet fermions but with its left- and right-handed components transforming in the same way under the gauge group $SU(2)\otimes U(1)$. 
These new particles have been searched for at hadron colliders, where they can be produced both singly and in pairs, with subsequent decays into a SM quark and a gauge boson or Higgs boson~\cite{delAguila:2000aa,delAguila:2000rc,AguilarSaavedra:2009es,Chala:2017xgc,Zhou:2019alr,Han:2014qia,Liu:2015kmo,Yang:2018oek,Liu:2019jgp,Yang:2010yh}. 
Searches at the LHC for VLTs have been performed and presented by the ATLAS and CMS Collaborations with the lower mass bounds on $T$ reaching up to about $740\sim1370$\,GeV, depending on the SU(2) multiplets they belong to and different branching fractions assumed~\cite{CMSbounds,ATLASbounds}.

Another motivation for BSM is the observation of neutrino oscillation in solar, atmospheric, reactor and accelerator experiments, which implies that neutrinos of three flavors are mixed and have tiny masses\,($\sim$ sub-eV)~\cite{Tanabashi:2018oca}. Various schemes have been proposed to include neutrino masses in the SM, among which the most popular one is the so-called seesaw mechanism~\cite{minkowski1977,yanagida1979,ms1980,sv1980,Weinberg:1979sa,mw1980,cl1980,lsw1981,ms1981,flhj1989,ma1998}, since it not only generates neutrino mass in an elegant way, but also connects the observed baryon asymmetry in the universe (BAU) and the origin of neutrino mass through leptogenesis~\cite{fy1986,krs1985,lpy1986,luty1992,mz1992,fps1995,crv1996,pilaftsis1997,ms1998,hs2004,Gu:2019nhb}. Some variations of the seesaw mechanism can also accommodate dark matter candidates~\cite{ma2006,ma2015,Gu:2019ogb,Zhou:2016jyp}. Among several seesaw mechanisms, Type-I seesaw~\cite{minkowski1977,yanagida1979,ms1980,sv1980} in a straightforward way introduces three singlet right-handed (RH) neutrinos ($N_{R\alpha}$), leading to Dirac mass terms as well as the RH Majorana mass terms. Consideration of both mass terms can generate sub-eV Majorana neutrino masses if the RH Majorana mass is set at $\sim10^{14}$\,GeV. 
Neutrino mass generation via seesaw mechanism comes with violation of lepton number (LNV) by $\Delta L=2$, which may be used on experiments as a sign for the Majorana nature of neutrinos. Many experiments and tests on the nature and properties of neutrinos are already in progress, such as the search for neutrinoless double beta decay ($0\nu\beta\beta$)~\cite{Furry:1939qr,Doi:1985dx}, as well as other LNV processes including rare $\tau$ decays, meson decays and hyperon decays~\cite{Ng:1978ij,Abad:1984gh,Dib:2000wm,Ali:2001gsa,Littenberg:1991rd,Barbero:2002wm}. However, the $\Delta L=2$ processes are suppressed either by a factor of $m^{2}_{\nu}/m^{2}_{W}$ due to the smallness of the light neutrino mass $m_{\nu}$, or by a factor of $|V_{\alpha m}V_{\beta m}|^{2}$ due to the small mixings, depending on whether the exchanged neutrino is light or heavy compared to the scale of the LNV processes~\cite{Atre:2009rg}. Fortunately, the LNV processes may be enhanced substantially as a result of resonant production of heavy neutrinos, if the heavy neutrino mass can be kinematically accessible (below TeV) as in some low-scale Type-I seesaw scenarios~\cite{Asaka:2005an,Asaka:2005pn,Asaka:2006nq,Asaka:2006ek,Xing:2009in,Adhikari:2010yt,Ibarra:2010xw,Boucenna:2014zba,Zhou:2017lrt,Gu:2018kmv}, which may be produced directly at collider experiments and searched for~\cite{deGouvea:2006gz,deGouvea:2007hks,Atre:2009rg,Kersten:2007vk,Bajc:2007zf,He:2009ua,Han:2006ip,Ibarra:2011xn,Dev:2013wba,Deppisch:2015qwa,Das:2018hph,Liu:2019qfa}. An upper bounds has been given by the LEP experiments on the mixings $|V_{eN,\mu N}|^{2}<\mathcal{O}(10^{-5})$ for heavy neutrino mass of $(80, 205)$\,GeV~\cite{Abreu:1996pa}. CMS broadened the mass range to $(20, 1600)$\,GeV and put a similar limit as $|V_{eN,\mu N}|^{2}<\mathcal{O}(10^{-5})$~\cite{Sirunyan:2018xiv,Sirunyan:2018mtv}.

To solve the above two problems of naturalness and neutrino oscillation, models have been proposed to incorporate neutrino masses into some scenarios with the VLTs, such as ones that include lepton-number violation between scalar triplet and lepton doublet within the Littlest Higgs scenario~\cite{Han:2005nk}, as well as other Little Higgs~\cite{delAguila:2019mvp,Dey:2008dk,deAlmeida:2007khx,Li:2011ao,Hektor:2007uu,Goyal:2006yn,Abada:2005rt,Goyal:2005it,delAguila:2005yi,Lee:2005mba,Chang:2003vs}, Composite Higgs~\cite{Coito:2019wte,Shindou:2017bem,Smetana:2013hm,delAguila:2010vg,Lee:2005kd}, Top Seesaw~\cite{He:1999vp,He:2001fz,Wang:2013jwa}, Higgs Inflation models~\cite{He:2014ora}, etc~\cite{Du:2012vh,Abe:2012fb}. A common feature of these new models is that VLTs and heavy Majorana neutrinos are predicted, which leads to a new decay mode of VLT through the mediation of heavy Majorana neutrinos. Taking into account that VLTs and heavy Majorana neutrinos in low-energy Type-I seesaw models are both within the search abilities at the LHC, in a model-independent way, we propose in this paper a search strategy for the above new decay channel of the VLT. There are some traditional ways to search  for the vector-like top partner at the LHC, such as $T \to b\,jj$. These searches have been performed~\cite{CMSbounds,ATLASbounds} and exclude the top partner mass up to 740$\sim$1370 GeV. In this work, we will investigate the process $T\to b\,\ell^{+}\ell^{+}jj$, which can be used to probe the top partner and to test the seesaw mechanism simultaneously at the LHC. In a scenario that incorporates three right-handed Majorana neutrinos and a top partner $T$ ($+2/3$ electrically charged and SU(2) singlet), we demonstrate that with the heavy Majorana neutrinos at GeV scale, the new decay mode of $T$ can be probed at the LHC by searching for a signal of same-sign dileptons~\cite{Cao:2011ew}, which has also been utilized in phenomenological study on topcolor-assisted technicolor model~\cite{Liu:2010ag} and rare B decay~\cite{Bao:2012vq}. In Section II we will introduce the relative effective Lagrangian of the scenario and then in Section III we will present our analysis by Monte-Carlo simulation of the search at the LHC and show the observability for the mixing $V_{Tb}$ between top partner and the SM quark, as well as the light-heavy neutrino mixing $V_{\mu N}$. Section IV is the conclusion.

\section{Effective Lagrangian and the new decay mode of top partner}
\label{sec:model}
\begin{figure}
\centering
\includegraphics[scale=0.34]{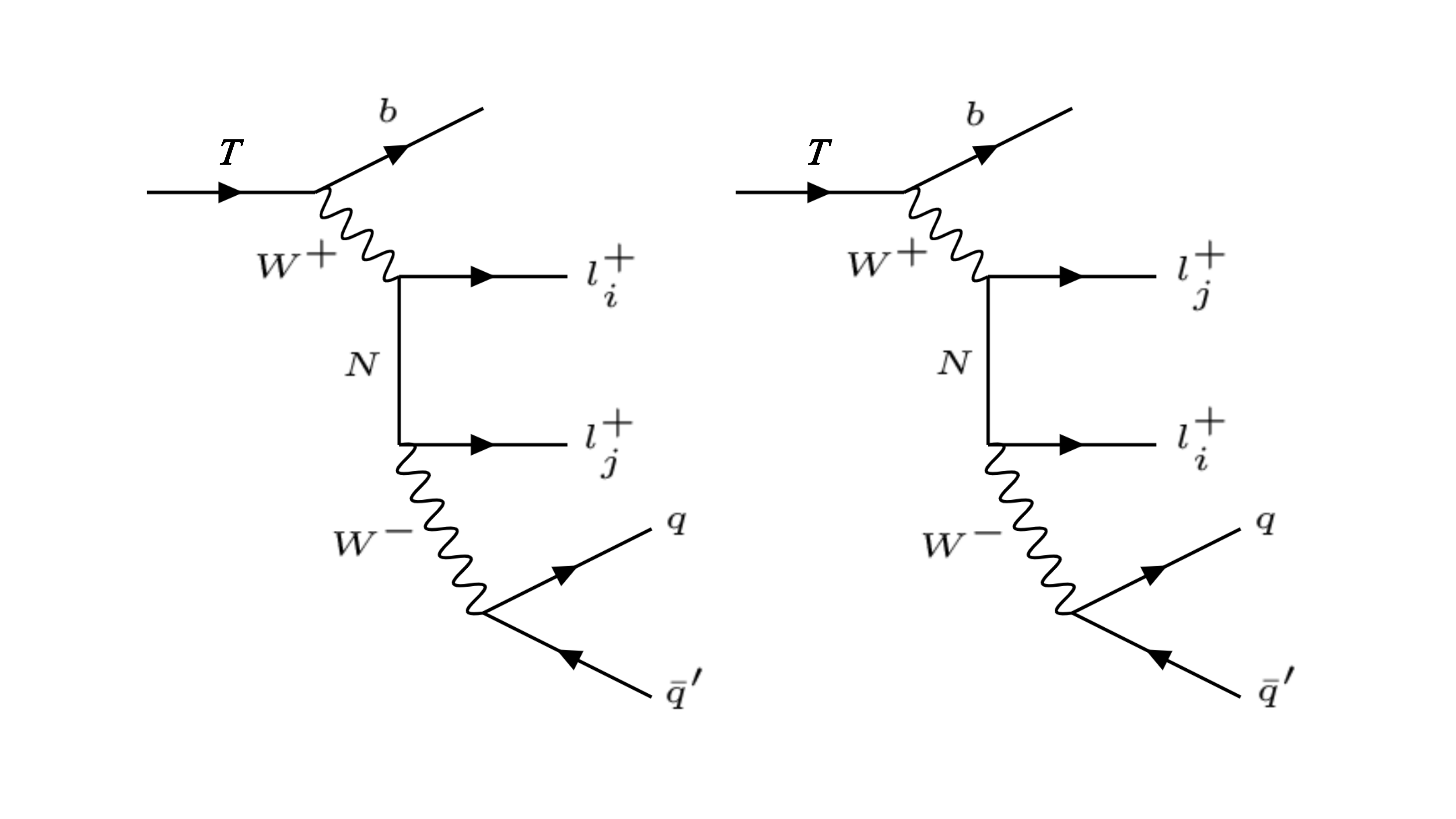}
\caption{Feynman diagrams of the top partner decay $T\to b\,\ell^{+}\ell^{+}q\bar{q'}$ including t- and u-channels.}
\label{fig:fd}
\end{figure}
Without losing generality, we will adopt the effective interaction method to perform a phenomenological study in the following sections. The relevant interaction between singlet top partner and $W$ boson is given by, 
\begin{align}
\mathcal{L}_{T}={}&-\frac{g}{\sqrt{2}}\bar{u}_{\alpha L}\gamma^{\mu}V_{\alpha i}d_{i L}W^{+}_{\mu}+\textrm{H.c.}\,,
\label{LT}
\end{align}
where $V$ is the CKM matrix but generalized to $4\times3$ dimensionally to include the additional VLT. Note that $\alpha=1\sim4$ runs over all generations of quarks including the VLT (here for brevity we label the singlet top partner $T$ as the 4th generation), while $i=1\sim3$ is the index for three generations of the SM fermions. $g$ is the weak coupling. As for the Type-I seesaw, for simplicity and without losing the features of low-scale Type-I seesaw, the model is parameterized as a mass scale of the RH Majorana neutrino ${M_N}$ and a mixing parameter between the light and heavy neutrinos $V_{ij}$, then the effective Lagrangian for interactions between the heavy Majorana neutrinos and charged leptons can be written as
\begin{align}
\mathcal{L}_{N}={}&-\frac{g}{2\sqrt{2}}V_{\ell N}W^{+}_{\mu}\ell\gamma^{\mu}(1-\gamma_{5})N^{c}+\textrm{H.c.}\,,
\label{LN}
\end{align}
where $\ell$\,($\ell=e,\mu,\tau$ are charged leptons) and $N$\,($N=N_{1,2,3}$ label heavy Majorana neutrinos) are mass eigenstates.

\begin{figure}
\centering
\includegraphics[scale=0.8]{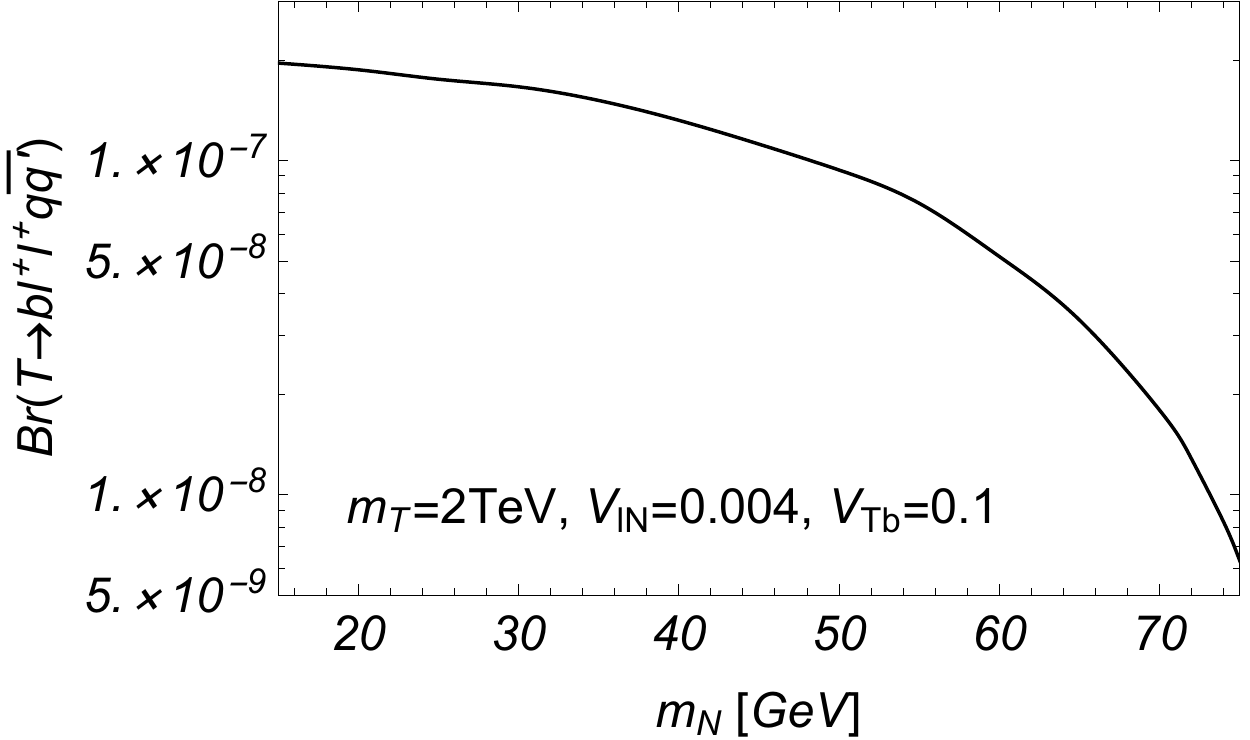}
\caption{Branching ratio of the top partner decay $T\to b\,\ell^{+}\ell^{+}q\bar{q'}$, with respect to the heavy Majorana neutrino mass. $\text{Br}(T\to bW^{+})=50\%$ is assumed.}
\label{fig:br}
\end{figure}
Introduction of interaction terms Eq.\eqref{LT} and \eqref{LN} leads to $T$ decay channel through the heavy Majorana neutrinos into a same-sign dilepton plus jets,
\begin{align}
T\to b\,W^{+}\to b\,\ell^{+}\ell^{+}q\bar{q'}\,,
\label{Tdecay}
\end{align}
the Feynman diagram of which is presented in \figurename~\ref{fig:fd}. The final same-sign dilepton may serve as a distinct signature at the LHC as a probe for this rare decay, as we will show in the next section. The relevant interaction terms for the process can be parameterized and written as an effective Lagrangian
\begin{align}
\mathcal{L}={}&-\frac{g}{2\sqrt{2}}W^{+}_{\mu}\left[V_{\ell N}\ell\gamma^{\mu}(1-\gamma_{5})N^{c}+V_{Tb}\bar{T}\gamma^{\mu}(1-\gamma_{5})b\right]+\textrm{H.c.}\,,
\label{Ltotal}
\end{align}
in which the indices are the same as above Eq.\eqref{LN}. Assuming $\text{Br}(T\to bW^{+})=50\%$, the branching ratio of the rare decay of $T$ is shown in \figurename~\ref{fig:br}, with respect to mass of the heavy Majorana neutrino. In the calculation we set $m_{T}=2$\,TeV, $V_{Tb}=0.1$ and $V_{\ell N}=0.004$ which are not excluded by current experiments (Note that $V_{\mu N}=0.004$ survives~\cite{Abreu:1996pa} in the mass range of \figurename~\ref{fig:br} while $V_{eN}$ does not~\cite{Agostini:2018tnm}). As can be seen, the branching ratio of the rare decay decreases as $m_{N}$ grows in the kinematically accessible range ($15\sim75$\,GeV). For a larger $m_{N}$ from $10^{2}$\,GeV to TeV scale, the rare decay can be enhanced a bit due to on-shell $W$ boson from $N$ decay, but the branching ratio of $T$ rare decay is still lower ($\sim10^{-8}$) than that in a light mass region. Therefore in Section III we will focus on the relatively small mass range of the heavy Majorana neutrino with $m_{N}\lesssim m_{W}$ and explore the possibilities for a probe of the top partner's new decay channel.

\section{Search at the LHC}
\label{sec:search}

In the scenario we introduced in the last section, the $SU(2)$ singlet top partner can be produced in pair via QCD processes at the LHC, and then goes through a rare decay mediated by the heavy Majorana neutrino. If one of the paired VLTs goes through the rare decay $T\to b\,\ell^{+}\ell^{+}jj$ and the other decays into hadrons $\bar{T}\to\bar{b}\,W^{-}\to b\,jj$, then we have the distinct signal at the hadronic environment of the LHC as a same-sign dilepton with multi-jets including two b-tagged ones:
\begin{align}
pp\to T\bar{T}\to\ell^{+}\ell^{+}+b\,\bar{b}+\,j_{1}j_{2}j_{3}j_{4}\,.
\label{signal1}
\end{align}
However, the GERDA experiments~\cite{Agostini:2018tnm} already put a very stringent limit on the $e$-flavor mixing $|V_{eN}|^{2}$ to about $10^{-8}$ in the GeV$\sim$100 GeV mass range of the heavy Majorana neutrino. Besides, probing for $\tau$-flavor mixing $V_{\tau N}$ requires accurate tagging of final $\tau$'s, which is not realized with high efficiency in the current collider simulation. Given these facts, we expect an improvement of sensitivity for the $\mu$-flavor mixing $V_{\mu N}$ in the kinematically accessible mass range of heavy neutrino ($m_{N}\lesssim m_{W}$ in our case), which can lead to its resonant production as we discussed in Section~\ref{sec:intro}. The contribution of CP-conjugate process of \eqref{signal1},
\begin{align}
pp\to T\bar{T}\to\ell^{-}\ell^{-}+b\,\bar{b}+\,j_{1}j_{2}j_{3}j_{4}
\label{signal2}
\end{align}
is also included in the following simulations. Note that the top partner can also be produced singly via electroweak processes. The cross section of the singly production is smaller than that of the pair production in a small mass range of VLT. As the VLT mass increases ($\gtrsim1.1$\,TeV), the singly production cross section will surpass that of pair production and lead to a better sensitivity for the new decay mode.

For the signal processes \eqref{signal1} and \eqref{signal2} whose final states contain a same-sign dilepton and jets, major SM backgrounds at the LHC come from prompt multi-leptons (mainly from events with $t\bar{t}\,W^{\pm}$ and $W^{\pm}W^{\pm}$+jets) and fake leptons (mainly from events with jets of heavy flavor, such as $t\bar{t}$). To be exact, opposite-sign dileptons with one of which mismeasured should also constitute our backgrounds, but as the rate of mismeasurement for muons is generally low enough that we ignore its effects, then the SM backgrounds we consider are
\begin{align}
pp\to{}& t\bar{t}\,W^{\pm},\notag\\
pp\to{}& W^{\pm}W^{\pm}+\text{jets},\notag\\
pp\to{}& t\bar{t}\,.
\label{bkg}
\end{align}
It should be noted that, in our scenario with heavy Majorana neutrinos introduced in the SM, the rare decay of top quark mediated by the heavy Majorana neutrinos:
\begin{align}
t/\bar{t}\to b/\bar{b}+\ell^{\pm}\ell^{\pm}+\text{jets}\,,
\end{align}
will also be present and contribute to the backgrounds $t\bar{t}$ and $t\bar{t}W^{\pm}$. We include these events as well in our simulation for the above backgrounds.

Monte-Carlo simulations are performed for both the signal \eqref{signal1}, \eqref{signal2} and backgrounds \eqref{bkg} at the LHC with the center-of-mass energy of 14 TeV. For the signal we specify the $N_{2}$-mediated decay processes for the probing of $V_{\mu N}$. As we adopt a diagonalized mixing matrix $V_{\ell N}$, thus $N_{2}$ couples only to the $\mu$-flavor. In the simulation, we use the benchmark point as following,
\begin{align}
m_{N}=50\,\text{GeV},\quad m_{T}=2\,\text{TeV},\quad V_{Tb}=0.1,\quad V_{\mu N}=1.0,
\end{align}
where by $m_{N}$ it means the mass of $N_{2}$ for simplicity, while for $N_{1}$ ($N_{3}$) that couples to e ($\tau$), we assume a kinematically inaccessible mass 300 GeV (1 TeV). Hence the processes mediated by $N_{1,3}$ are not included in the simulation of the search for the VLT new decay, due to their inaccessible large masses. Parton-level events of signal and backgrounds are generated through \textsf{MadGraph5\_aMC@NLO}~\cite{Alwall:2014hca} with the NN23LO1 PDF~\cite{Ball:2012cx}, then go through parton showering and hadronization by \textsf{Pythia-8.2}\,\cite{pythia}. The renormalization and factorization scales are set at the value of VLT mass, that is, 2000 GeV. Detector simulations are carried out by tuned \textsf{Delphes3}~\cite{delphes} within the framework of \textsf{CheckMATE2}~\cite{Dercks:2016npn}. Jet-clustering is done using \textsf{FastJet}~\cite{fastjet} with anti-$k_t$ algorithm~\cite{anti-kt}. We assume b-tagging efficiency to be $70\%$ in our simulation. In addition, contributions from higher order QCD corrections are taken into account by normalizing the leading-order cross sections of $t\bar{t}$ and $t\bar{t}\,W^\pm$ to NNLO and NLO, respectively~\cite{Czakon:2011xx,Frixione:2015zaa}.

\begin{figure}[t]
\centering
\begin{minipage}{0.32\linewidth}
  \centerline{\includegraphics[scale=0.34]{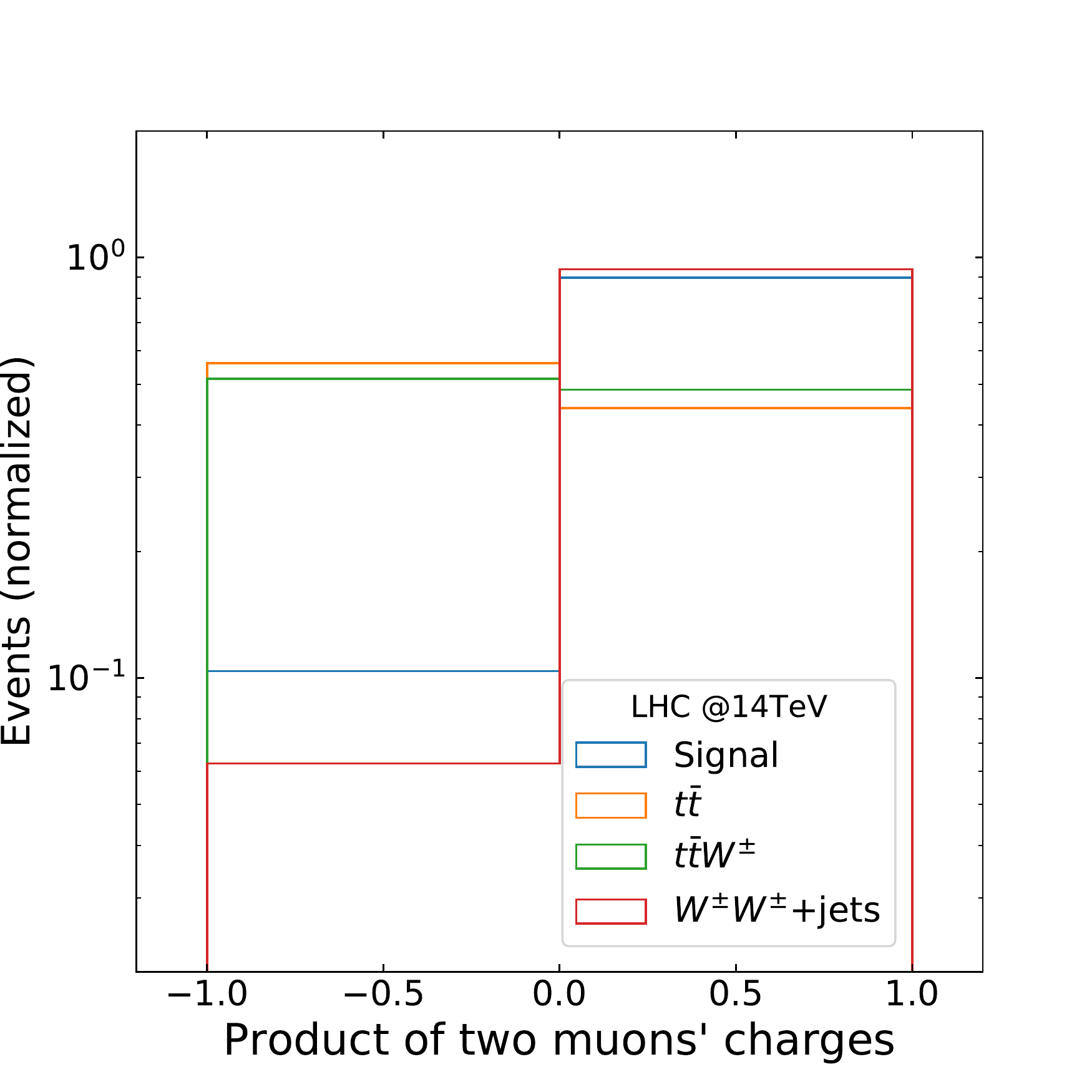}}
  \centerline{(a)}
  \label{fig:cc}
\end{minipage}
\hfill
\begin{minipage}{0.32\linewidth}
  \centerline{\includegraphics[scale=0.34]{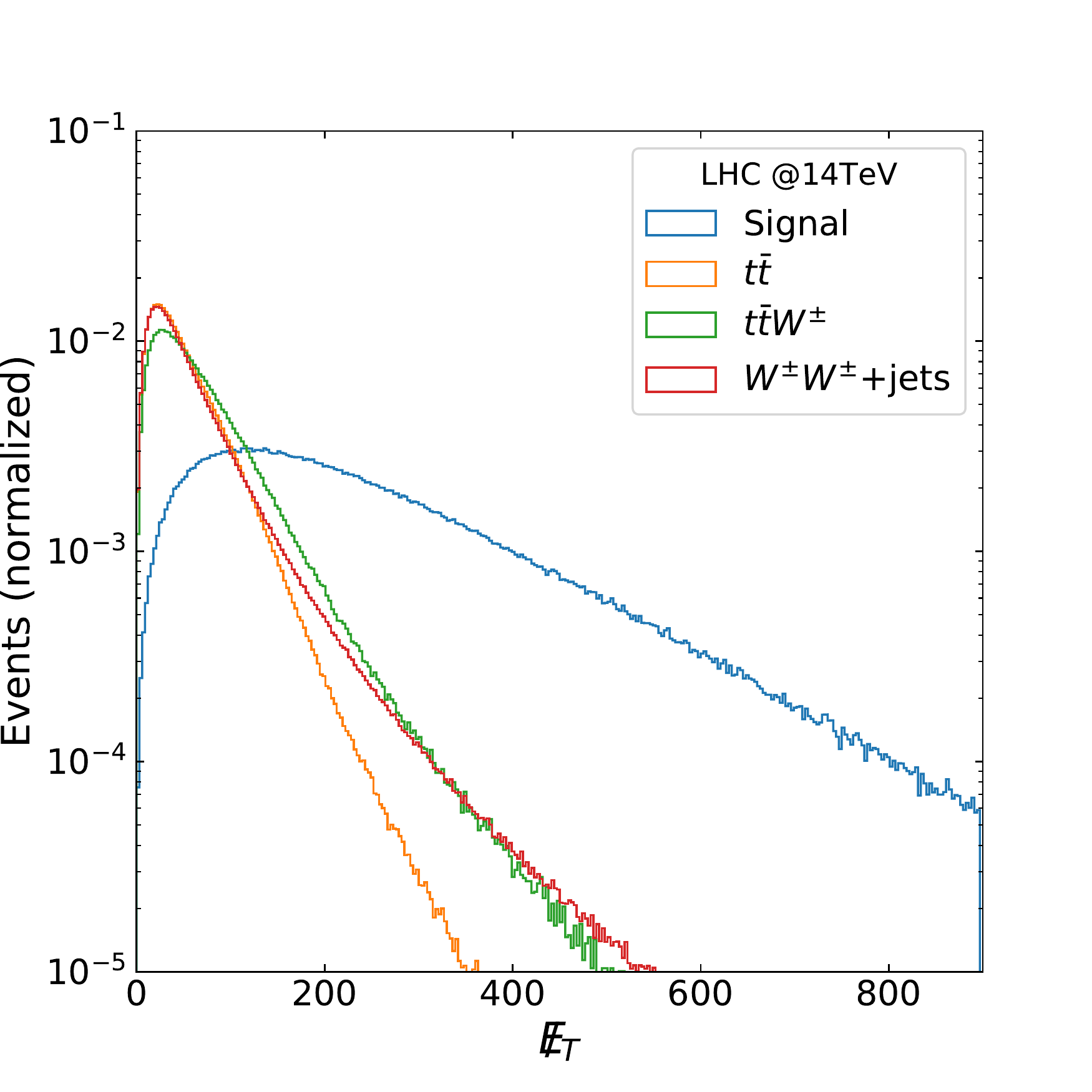}}
  \centerline{(b)}
\end{minipage}
\hfill
\begin{minipage}{0.32\linewidth}
  \centerline{\includegraphics[scale=0.34]{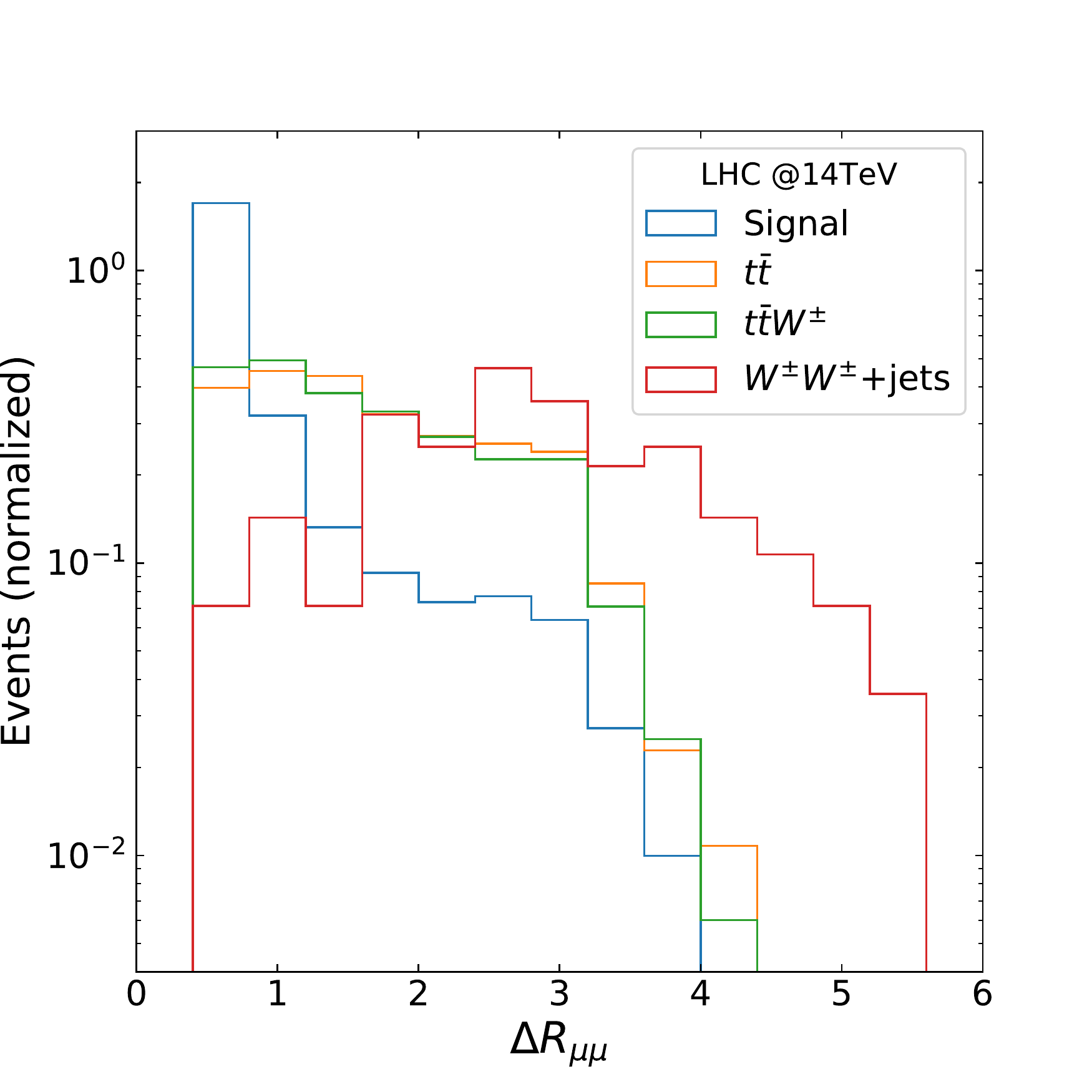}}
  \centerline{(c)}
\end{minipage}
\\
\begin{minipage}{0.49\linewidth}
  \centerline{\includegraphics[scale=0.34]{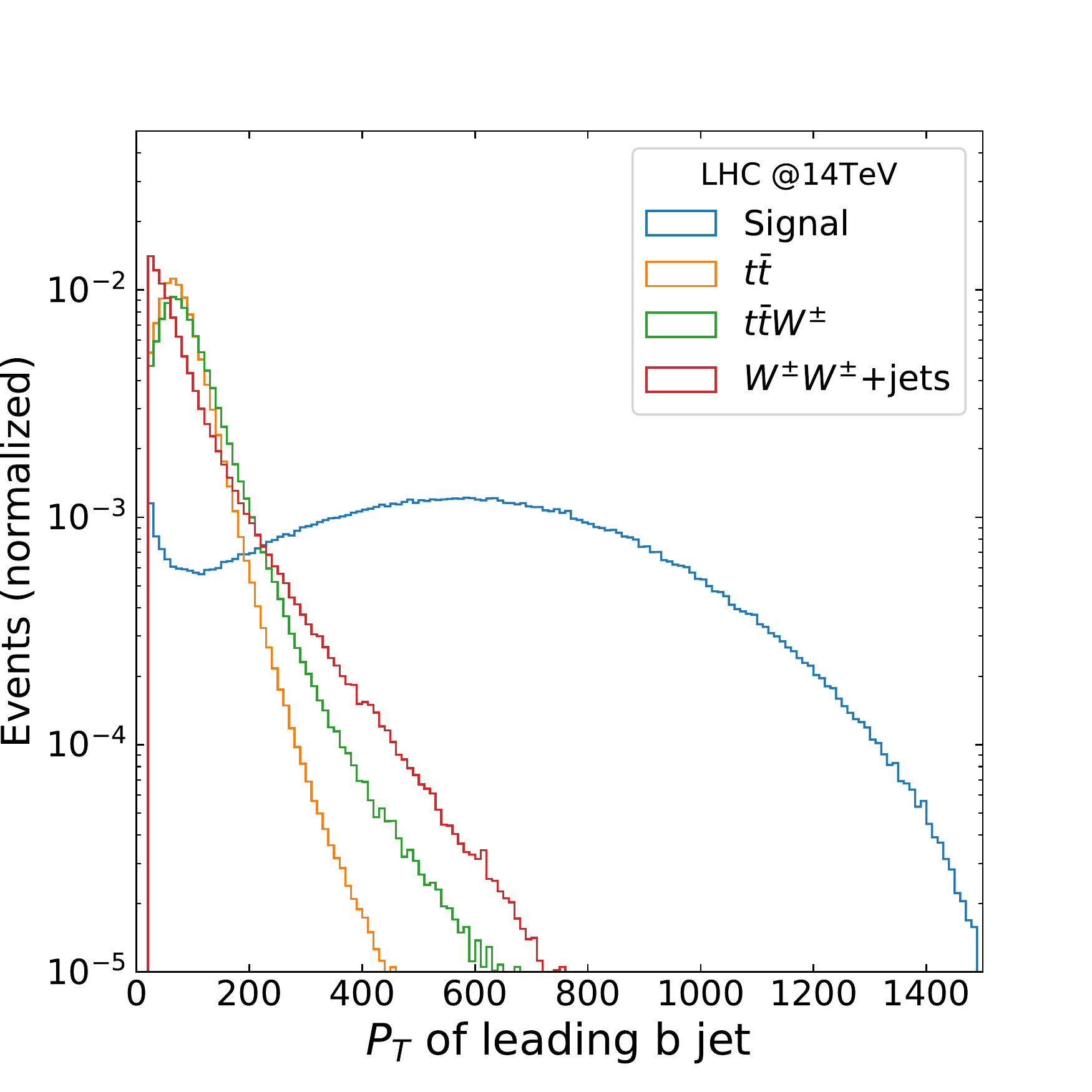}}
  \centerline{(d)}
\end{minipage}
\hfill
\begin{minipage}{0.49\linewidth}
  \centerline{\includegraphics[scale=0.34]{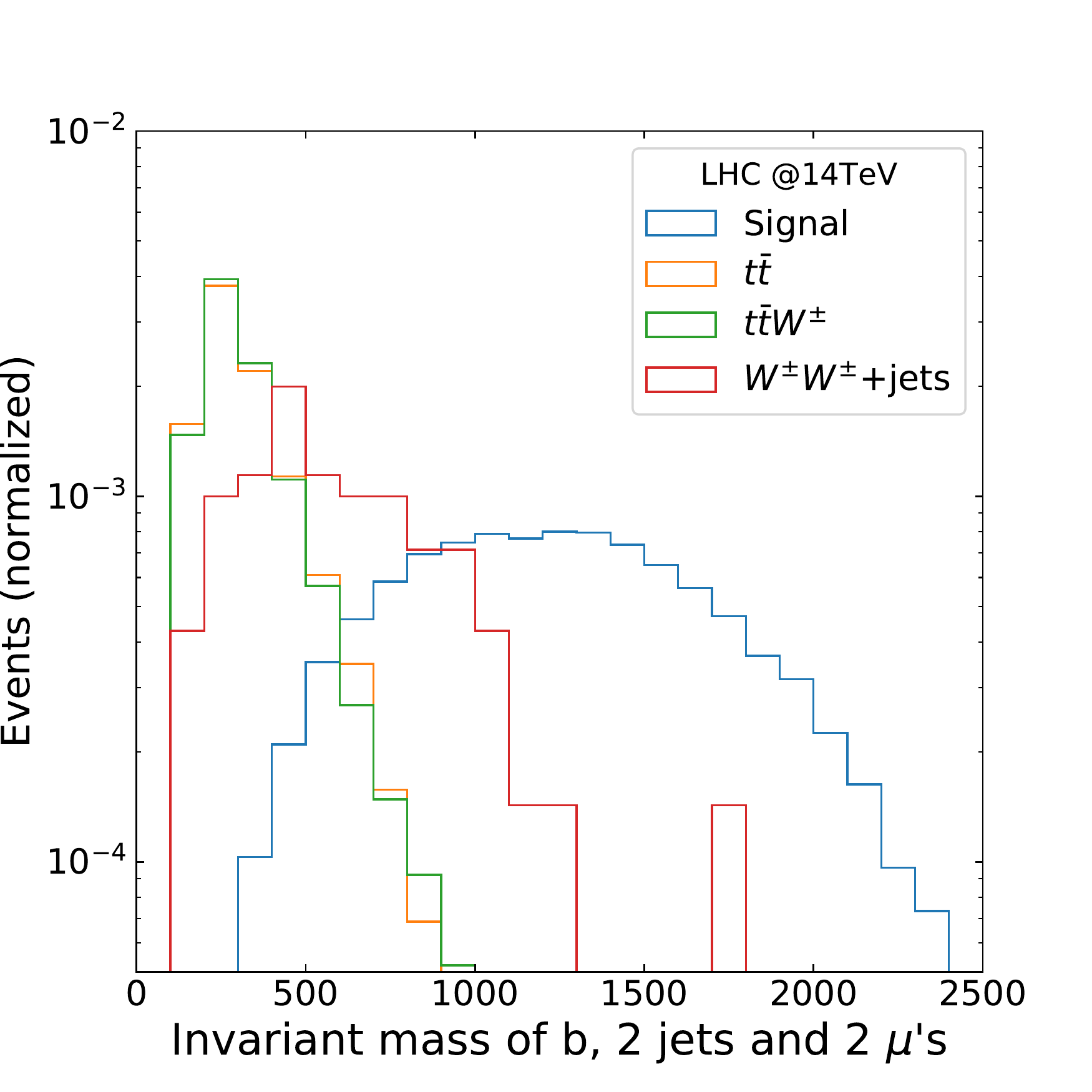}}
  \centerline{(e)}
\end{minipage}
\caption{Kinematic distributions of the signal $pp \to \mu^{\pm}\mu^{\pm}+2b+4j$ and the SM backgrounds $t\bar{t}$, $t\bar{t}\,W^\pm$, $W^\pm W^\pm$+jets at the 14 TeV LHC. The benchmark point is chosen as $m_{T}=2$\,TeV, $m_{N}=50$\,GeV, $V_{Tb}=0.1$, $V_{\mu N}=1.0$.}
\label{fig:dist}
\end{figure}

We present the distributions of kinematic variables for signal and the SM background processes at the 14\,TeV LHC in \figurename~\ref{fig:dist}, including the product of charges of the final dimuon (\figurename~\ref{fig:dist}-a), missing transverse energy $\slashed{E}_{T}$ (\figurename~\ref{fig:dist}-b), the relative distance between the final dimuon $\Delta R_{\mu\mu}$ (\figurename~\ref{fig:dist}-c) and the transverse momentum of leading b-jet (\figurename~\ref{fig:dist}-d). \figurename~\ref{fig:dist}-e is the distribution of $m_{b2\mu2j}$, the invariant mass reconstructed from the final two muons, a leading b-jet and two jets. It can be seen from (\figurename~\ref{fig:dist}-a) that the charges of final dimuon for the backgrounds $t\bar{t}$ and $t\bar{t}\,W^{\pm}$ tend to be opposite, compared with the signal. The distribution of missing transverse energy for the signal is more flat than that for the backgrounds and more signal events are found in the range of a large $\slashed{E}_{T}$ ($100\,\text{GeV}\sim900$\,GeV). Intuitively there are no neutrinos present in the final states of signal and hence signal events tend to have smaller $\slashed{E}_{T}$ compared with the backgrounds, in which neutrinos from $W$ decay constitute much of the missing transverse energy. However, note that it is the jets after parton-showering that arrive at the detectors, rather than the partonic final states. $b$\textbackslash$\bar{b}$ quarks from the signal \eqref{signal1} and \ref{signal2} are highly energetic as they are decay products of $T$ with mass of 2\,TeV and by parton-showering, these $b$\textbackslash$\bar{b}$ quarks decay to neutrinos that are highly energetic as well, leading to the $\slashed{E}_{T}$ distribution shown in \figurename~\ref{fig:dist}-b. The dimuon in the signal comes from decay of the same top partner (Eq.\eqref{signal1} and \eqref{signal2}) while in the SM backgrounds, the final leptons come from decays of different parent particles, that is, $W^{+}W^{-}$ in the $t\bar{t}$, $t\bar{t}\,W^{\pm}$ and $W^{\pm}W^{\pm}$+jets events. The final two muons thus tend to be closer in the signal than in the backgrounds, which is reflected in the distributions of the relative distance between them (\figurename~\ref{fig:dist}-c). Moreover, we set $m_{T}$ in the benchmark point to be 2\,TeV, which is much more massive than the SM top quark and whose decay product b quark tends to be much harder than that from top decay in the backgrounds event (\figurename~\ref{fig:dist}-d). Furthermore, to distinguish the signal and backgrounds more efficiently, we reconstruct the parent $T$ by the invariant mass $m_{b2\mu2j}$ clustering the decay products from the VLT rare decay, in which we use the leading b-jet and two soft jets, since the jets come from the secondary decay of the mediating Majorana neutrino and hence tend to be softer than ones in the SM backgrounds. As can be seen from \figurename~\ref{fig:dist}-e, more events of the signal distribute around the range $800\,\text{GeV}\lesssim m_{b2\mu2j}\lesssim 2000$\,GeV, while for the three backgrounds the peaks of the distributions are all below 800\,GeV.
\begin{table}
\centering
\begin{tabular}{l|c|c|c|c}
\hline\hline
 & $t\bar{t}$ & $t\bar{t}\,W^{\pm}$ & $WW+$jets & signal \\ \hline
Cut 1: Same-sign dimuon & 16.30 & $2.84\times10^{-2}$ & $1.77\times10^{-3}$ & 0.590  \\ \hline
Cut 2: At least 6 jets & 4.06 & $9.20\times10^{-3}$ & $1.10\times10^{-4}$ & 0.268  \\ \hline
Cut 3: $\slashed{E}_{T}>180$\,GeV & $4.30\times10^{-2}$ & $5.88\times10^{-4}$ & $2.24\times10^{-5}$ & 0.116  \\ \hline
Cut 4: On relative distance\, & $1.15\times10^{-2}$ & $1.55\times10^{-4}$ & $1.74\times10^{-6}$ & $9.22\times10^{-2}$  \\ \hline
Cut 5: At least 1 b-jet\,($p_{T}>280$\,GeV) & $7.87\times10^{-4}$ & $1.02\times10^{-5}$ & $0$ & $4.87\times10^{-2}$  \\ \hline
Cut 6: $m_{b2\mu2j}>1200$\,GeV & $3.48\times10^{-5}$ & $3.35\times10^{-6}$ & $0$ & $2.71\times10^{-2}$ \\
\hline\hline
\end{tabular}
\caption{Cutflow of cross sections for the signal process $pp \to \mu^{\pm}\mu^{\pm}+2b+4j$ and the SM background processes $pp \to t\bar{t}, t\bar{t}\,W^\pm, WW$+jets at the 14 TeV LHC. The cross sections are in the unit of pb. The benchmark point is the same as in \figurename~\ref{fig:dist}.}
\label{tab:cutflow}
\end{table}
Based on the above analysis, we apply the following kinematic cuts to the events to distinguish signal from the SM backgrounds.
\begin{itemize}
\item Cut 1: A same-sign muons is required with each of them satisfying $p_{T}(\mu)>10$\,GeV and $|\eta(\mu)|<2.8$.
\item Cut 2: We demand at least 6 jets in the final states with each of them has $p_{T}(j)>15$\,GeV and $|\eta(j)|<3.0$.
\item Cut 3: A large missing transverse energy is required as $\slashed{E}_{T}>180$\,GeV.
\item Cut 4: The relative distance is also required for the dilepton separation as $0.4<\Delta R_{\mu\mu}<1.2$, for jets separation as $\Delta R_{jj}>0.4$ and for jet-lepton separation as $\Delta R_{\mu j}>0.4$.
\item Cut 5: Among the final jets, at least one of them is required to be a b-jet. And the leading b-jet should have $p_{T}>280$\,GeV.
\item Cut 6: The invariant mass $m_{b2\mu2j}$ clustering the decay products of $T$ rare decay is required to be larger than 1200\,GeV.
\end{itemize}

With the above cuts applied, we present in \tablename~\ref{tab:cutflow} the cutflow of cross sections for signal and the SM backgrounds at the 14\,TeV LHC. Among the three kinds of SM backgrounds, we can see from \tablename~\ref{tab:cutflow} that the dominant one is the $t\bar{t}$ events. The first two cuts on numbers of final same-sign muons and jets can suppress the main backgrounds to the same order as the signal. Then the large $\slashed{E}_{T}$ requirement can cut about 95\% SM backgrounds while keeping 40\% signal events that have survived from the first two steps of cuts. The final three cuts can effectively remove the $WW$+jets events while leaving the $t\bar{t}$ and $t\bar{t}\,W^{\pm}$ backgrounds to an almost negligible level (about 3 orders smaller than the signal). Through the above event selection, we can reach an effective probe towards the parameter space in the present scenario.

\begin{figure}[t]
\centering
\begin{minipage}{0.48\linewidth}
  \centerline{\includegraphics[scale=0.65]{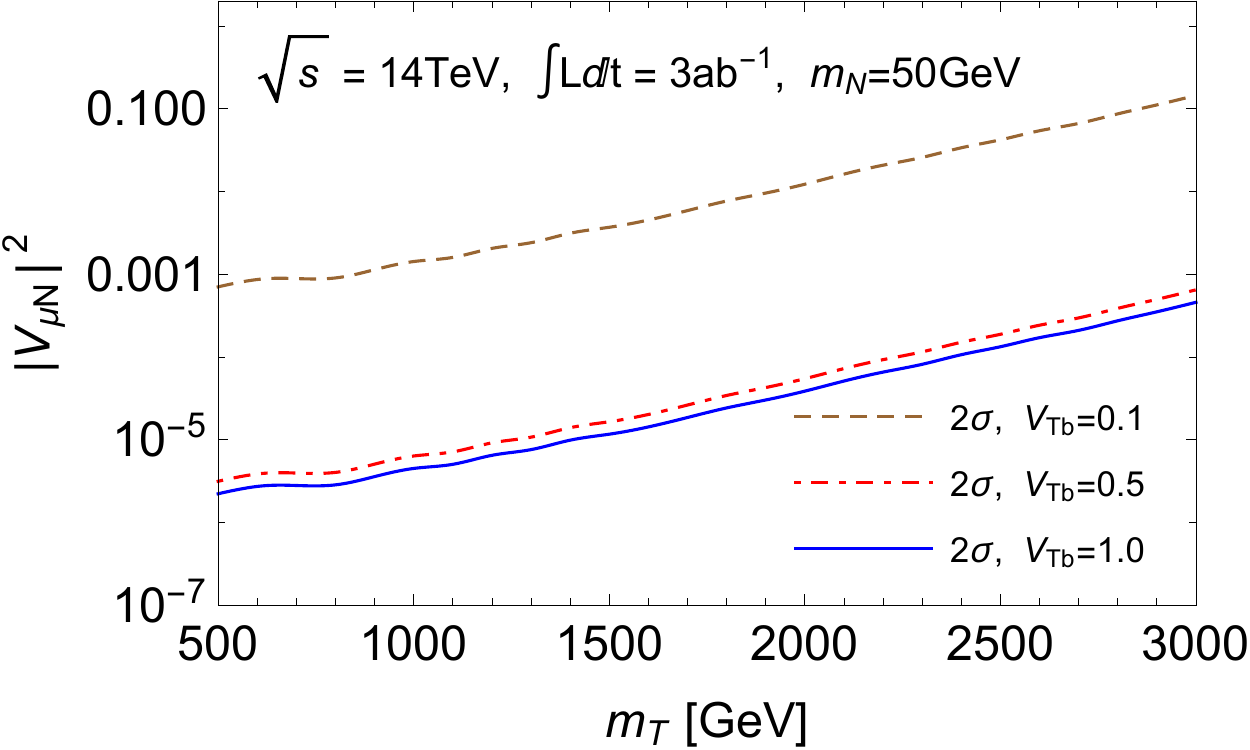}}
  \centerline{(a)}
\end{minipage}
\hfill
\begin{minipage}{0.48\linewidth}
  \centerline{\includegraphics[scale=0.65]{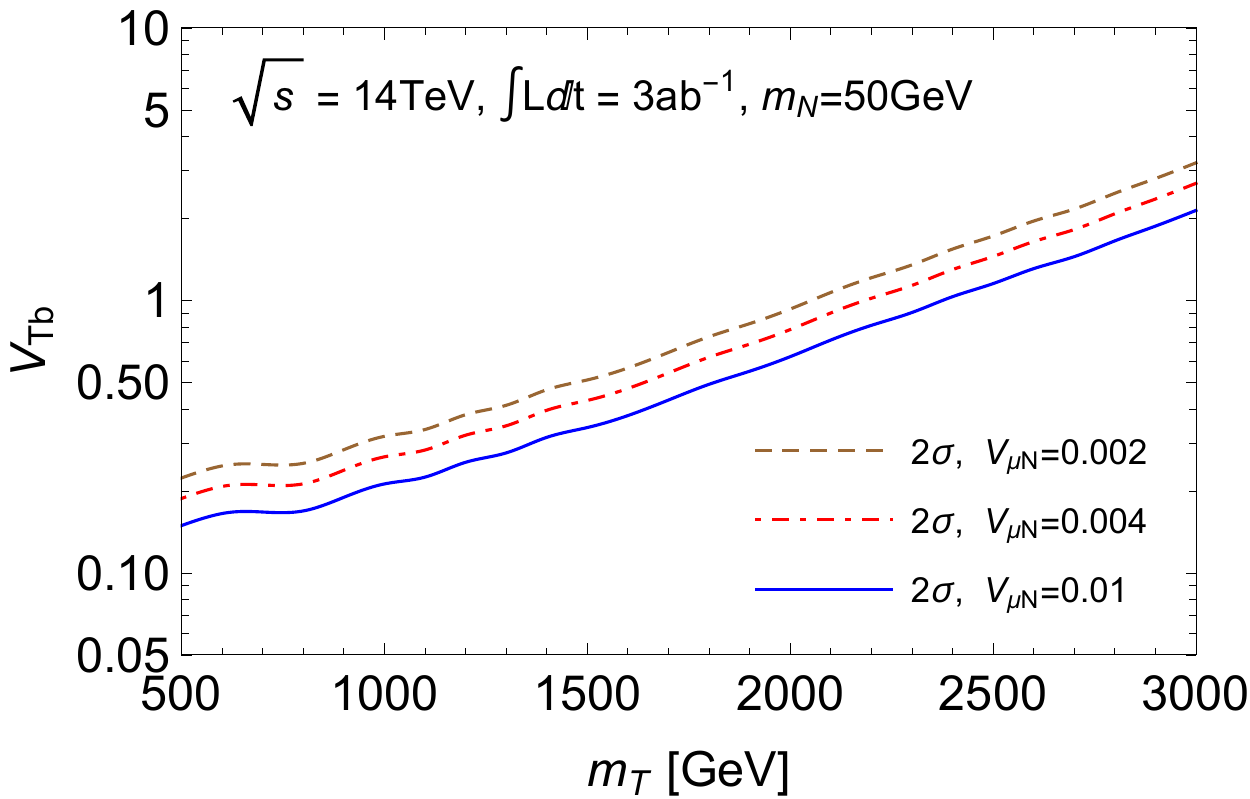}}
  \centerline{(b)}
\end{minipage}
\caption{Contours of $2\sigma$ exclusion limits on the signal $pp\to T\bar{T}\to\mu^{\pm}\mu^{\pm}+b\bar{b}+4j$ at the 14 TeV LHC, with integrated luminosity of $3\,\text{ab}^{-1}$. Systematic uncertainty $\beta$ is taken as 5\%. (a) is plotted on the plane of $|V_{\mu N}|^{2}$ versus $m_{T}$ and (b) on the plane of $V_{Tb}$ versus $m_{T}$.}
\label{fig:exclu}
\end{figure}

\begin{figure}[t]
\centering
\includegraphics[scale=0.7]{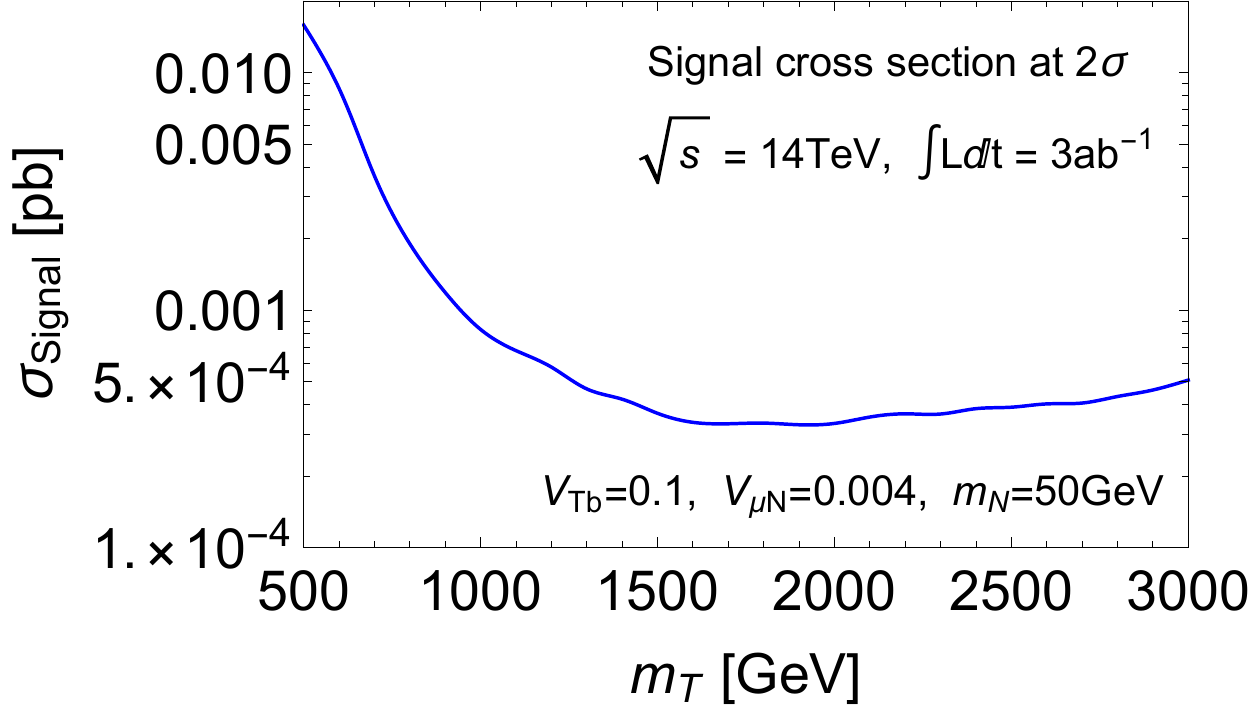}
\caption{$2\sigma$  exclusion on the cross sections of signal $pp\to T\bar{T}\to\mu^{\pm}\mu^{\pm}+b\bar{b}+4j$ with respect to the top partner mass at the 14\,TeV LHC.}
\label{fig:xsatsgnf2}
\end{figure}

We present in \figurename~\ref{fig:exclu} the exclusion bounds at $2\sigma$ on the signal $pp\to T\bar{T}\to\mu^{\pm}\mu^{\pm}+b\bar{b}+4j$, where the statistical significance is calculated using the formula
\begin{align}
\alpha=\frac{S}{\sqrt{B+(\beta B)^{2}}},
\end{align}
where $S$\,($B$) is the event number after the above cuts for signal\,(background). $\beta$ denotes the systematic uncertainty which in our case mainly comes from the background with misidentified leptons and is assumed to be 5\% in the following discussion. The integrated luminosity at the 14\,TeV LHC is set to be $3\,\text{ab}^{-1}$. \figurename~\ref{fig:exclu}\,(a) shows the $2\sigma$ limits on the parameter plane of $|V_{\mu N}|^{2}$ versus $m_{T}$, in which three colored lines correspond to $V_{Tb}=0.1,\,0.5,\,1.0$, respectively. It can be seen that for a relatively low mass range of $m_{T}\lesssim1.3$\,TeV and $V_{Tb}\gtrsim0.5$, the heavy-light Majorana neutrino mixing $|V_{\mu N}|^{2}$ can be probed to orders of $10^{-5}$ and below, which surpasses the current bounds on $|V_{\mu N}|^{2}\sim10^{-5}$ for $m_{N}\sim50$\,GeV given by the DELPHI Collaboration~\cite{Abreu:1996pa}, as well as the LHC search for same-sign dileptons~\cite{Sirunyan:2018xiv} and trileptons events~\cite{Sirunyan:2018mtv}. For example, we can read from \figurename~\ref{fig:exclu}\,(a) that, at the point of $m_{T}=1.3$\,TeV and $V_{Tb}=1.0$, the exclusion limit on $|V_{\mu N}|^{2}$ reaches $7.54\times10^{-6}$. \figurename~\ref{fig:exclu}\,(b) is plotted on the plane of $V_{Tb}$ versus $m_{T}$, in which three colored lines correspond to $V_{\mu N}=0.002,\,0.004,\,0.01$, respectively. Within the mass range of $m_{T}$ below 2.05\,TeV, the upper bound on the coupling of the top partner with the SM $b$ quark can be given at $V_{Tb}\lesssim1.0$ with $V_{\mu N}\sim10^{-3}$. For instance, at $V_{\mu N}=0.004$ and $m_{T}=1.3$\,TeV, $V_{Tb}>0.346$ can be excluded at $2\sigma$. Finally in \figurename~\ref{fig:xsatsgnf2} we present the $2\sigma$ exclusion limit on the cross sections of the signal process $pp\to T\bar{T}\to\mu^{\pm}\mu^{\pm}+b\bar{b}+4j$ with respect to the top partner mass at the 14\,TeV LHC, with $V_{Tb}=0.1$, $V_{\mu N}=0.004$ and $m_{N}=50$\,GeV. Furthermore, it can be inferred from \figurename~\ref{fig:br} that with a less massive heavy Majorana neutrino, the branching ratio of the top partner's rare decay will increase and may lead to a better sensitivity.

It should be noted that we have not considered pileup effects in our discussion, which is important for a fully realistic simulation and needs proper removal techniques~\cite{Cacciari:2007fd,Berta:2014eza,Krohn:2013lba}. However, the pileup effects can be limited on our results since the event-selection is based on a pair of hard same-sign leptons.

\section{Conclusion}
In this paper, we investigate the observability for the rare decay of a singlet top partner in a model-independent scenario that combines the low-energy Type-I seesaw and a vector-like singlet top partner. 
We present a search strategy at the 14\,TeV LHC for a distinguishable signal with a same-sign dilepton plus multiple jets. In a kinematically accessible mass range of the heavy Majorana neutrino (we choose $m_{N}=50$\,GeV as a benchmark point), the detector-level simulation at the 14\,TeV LHC with integrated luminosity of $3\,\text{ab}^{-1}$ shows that, the $\mu$-flavor mixing with the heavy Majorana neutrino $|V_{\mu N}|^{2}>7.54\times10^{-6}$ can be excluded at $2\sigma$ for $V_{Tb}\lesssim1.0$ and $m_{T}\sim1.3$\,TeV. The coupling between the singlet top partner and the SM $b$ quark $V_{Tb}>0.346$ can be excluded at $2\sigma$ for $V_{\mu N}=0.004$ and $m_{T}\sim1.3$\,TeV. It is then concluded that in a kinematically accessible mass range of the heavy Majorana neutrino, searching at the LHC for the rare decay of a singlet top partner mediated by the heavy Majorana neutrino can be promising through the same-sign dilepton signal.

\section{acknowledgments}
This work is supported by the National Natural Science Foundation of China (NNSFC) under grant Nos.\,11847208 and 11705093, as well as the Jiangsu Planned Projects for Postdoctoral Research Funds under grant No.\,2019K197.

\end{document}